\documentclass[11pt]{article}
\usepackage[titletoc,toc]{appendix}
\usepackage{amssymb}
\usepackage[numbers,sort&compress]{natbib}
\usepackage{amsmath,amsfonts,graphicx,epsfig}
\usepackage{bm}
\usepackage{xcolor}
\usepackage{color}
\usepackage{soul}
\def\bea{\begin{eqnarray}}
\def\eea{\end{eqnarray}}
\def\be{\begin{equation}}
\def\ee{\end{equation}}
\def\ba{\begin{array}}
\def\ea{\end{array}}
\def\nn{\nonumber}

\def\p{{\bf p}}

\def\k{{\bf k}}

\def\x{{\bf x}}
\def\y{{\bf y}}

\def\R{{\mathcal R}}
\def\S{{\mathcal S}}

\def\P{{\mathcal P}}

\def\X{{\mathcal X}}
\def\Y{{\mathcal Y}}
\def\N{{\mathcal N}}
\def\C{{\mathcal C}}

\begin{document}

\setlength\arraycolsep{2pt}

\renewcommand{\theequation}{\arabic{section}.\arabic{equation}}
\setcounter{page}{1}

\begin{titlepage}

\begin{flushright}
ICCUB-16-029
\end{flushright}

\begin{center}

\vskip 1.0 cm

{\LARGE  \bf Cumulative effects in inflation with ultra-light entropy modes}

\vskip 1.0cm

{\Large
Ana Ach\'ucarro$^{a, b}$, Vicente Atal$^{a}$, Cristiano Germani$^{c}$\\ and Gonzalo A. Palma$^{d}$
}

\vskip 0.5cm

{\it $^{a}$Instituut-Lorentz for Theoretical Physics, Universiteit Leiden, \mbox{2333 CA Leiden, The Netherlands.} \\
$^{b}$Department of Theoretical Physics, University of the Basque Country, \mbox{48080 Bilbao, Spain.}\\
$^{c}$ Institut de Ci\'encies del Cosmos, Universitat de Barcelona, Mart\'i i Franqu\'es 1, 08028 Barcelona, Spain.\\
$^{d}$Grupo de Cosmolog\'ia y Astrof\'isica Te\'orica, Departamento de F\'{i}sica, FCFM, \mbox{Universidad de Chile}, Blanco Encalada 2008, Santiago, Chile.
}

\vskip 1.5cm

\end{center}

\begin{abstract}

In multi-field inflation  one  or  more  non-adiabatic  modes  may  become light, potentially inducing large levels of isocurvature perturbations in the cosmic microwave background. If in addition these light modes are coupled to the adiabatic mode, they influence its evolution on super horizon scales. Here we consider the case in which a non-adiabatic mode becomes approximately massless (``ultralight") while still coupled to the adiabatic mode, a typical situation that arises with pseudo-Nambu-Goldstone bosons or moduli. This ultralight mode freezes on super-horizon scales and acts as a constant source for the curvature perturbation, making it grow linearly in time and effectively suppressing the isocurvature component. We identify a St\"uckelberg-like emergent shift symmetry that underlies this behavior. As inflation lasts for many $e$-folds, the integrated effect of this source enhances the power spectrum of the adiabatic mode, while keeping the non-adiabatic spectrum approximately untouched. In this case, towards the end of inflation all the fluctuations, adiabatic and non-adiabatic, are dominated by a single degree of freedom.

\end{abstract}

\end{titlepage}

\setcounter{equation}{0}
\section{Introduction}

Light scalar fields, with masses much smaller than the Hubble expansion rate $H$, are known to produce potentially large levels of entropy perturbations during inflation~\cite{Guth:1980zm, Starobinsky:1980te, Mukhanov:1981xt, Linde:1981mu, Albrecht:1982wi} that could lead to (1) non-adiabatic features in the Cosmic Microwave Background (CMB)~\cite{Langlois:1999dw, Amendola:2001ni, Bartolo:2001rt}, and (2) the super-horizon evolution of curvature perturbations~\cite{Gordon:2000hv, GrootNibbelink:2000vx, GrootNibbelink:2001qt, Wands:2002bn, Tsujikawa:2002qx, Byrnes:2006fr, Choi:2007su, Gao:2009qy}. 

In this article we study the extreme situation in which a non-adiabatic mode is approximately massless, and its interaction with the curvature perturbation persists during the whole period of inflation, from horizon crossing until reheating. In this case, the non-adiabatic mode freezes on super-horizon scales acting as a constant source for the growth of the adiabatic mode. Because inflation lasts for many $e$-folds, the adiabatic mode eventually becomes dominated by the particular solution sourced by the light mode,  which is a linearly growing function of time on super-horizon scales. By the end of inflation, the adiabatic mode may then be completely determined by the value of the light field on super-horizon scales. In this case, the associated power spectrum of curvature perturbations is enhanced by a factor proportional to the number of $e$-folds, squared. Interestingly however, the non-adiabatic perturbations do not experience the same cumulative effect, reducing the potential impact of isocurvature perturbations (relative to curvature perturbations) to the CMB~\cite{Planck:2013jfk, Ade:2015lrj}.

The mechanism discussed here, could be shared by many specific models involving light scalar fields such as pseudo-Nambu-Goldstone bosons and moduli. We will offer a concrete example where an ultra-light field emerges, that appears within a well studied class of models~\cite{GarciaBellido:1995fz, DiMarco:2002eb, DiMarco:2005nq, Lalak:2007vi, Cremonini:2010sv, Cremonini:2010ua, vandeBruck:2014ata} consisting of a multi-field action with a non-canonical kinetic term, typical of supergravity and string theory compactifications. As we will show, the appearance of the ultra-light field is related to a symmetry under field reparametrizations, mildly broken by slow-roll.

We emphasize that the phenomenology of models with ultra-light fields differ from curvaton scenarios~\cite{Lyth:2001nq, Lyth:2002my} where primordial fluctuations are generated at the very end of inflation. Here, instead, primordial fluctuations are generated during the whole process of inflation.

\setcounter{equation}{0}
\section{Coupling the adiabatic mode to an ultra-light field} \label{sec:coupling-R-sigma}

Let us start by recalling that in single field inflation, the quadratic action describing the dynamics of the adiabatic curvature perturbation $\R$ in a Friedmann-Robertson-Walker spacetime (in comoving gauge), is given by
\be
S_0=\int \! d^4x \, a^3 \epsilon \left[ \dot \R^2- \frac{1}{a^2} (\nabla \R)^2 \right],
\ee
where $\epsilon\equiv -\dot H/H^2$ and $H=\dot a/a$ (we work in units where the reduced Planck mass is set to unity $M_{\rm Pl} = 1$). In multi-field inflation, the generic situation is that there is at least one additional non-adiabatic perturbation $\sigma$. At quadratic order, the only allowed interaction compatible with diffeomorphism invariance of the system and the symmetries of the background, is an operator proportional to $\dot \R \sigma$ (note that, because we assumed that $\R$ already describes the adiabatic mode, a kinetic mixing of the form $\dot \R \dot\sigma$ is excluded~\cite{Langlois:2008mn}). Thus, the quadratic action (with unit sound speeds for both fields) coupling $\R$ and $\sigma$ is\footnote{For a complementary discussion on the construction of multi-field actions within the effective field theory approach of inflation~\cite{Cheung:2007st} see ref.~\cite{Senatore:2010wk}.}
\be\label{new-action-quadratic0}
S=\int \! d^4x \, a^3  \left[\epsilon \dot \R^2 - 2 \epsilon \alpha \dot \R \sigma  - \frac{ \epsilon }{ a^2 } \, (\nabla \R)^2 + \frac{1}{2}\left(\dot\sigma^2-\frac{1}{a^{2}}(\nabla\sigma)^2\right)-\frac{1}{2} m_\sigma^2\sigma^2\right] ,
\ee
where $m_\sigma$ is the so called effective mass of $\sigma$. Because of the background symmetries, $\alpha$ and $m_\sigma$ are time dependent parameters. The entropy perturbation $\sigma$ may be considered as massive, although one has to be careful about how to identify the relevant mass parameters that characterize the behavior of the perturbations.\footnote{More precisely, because of the non-diagonal coupling $ \dot \R \sigma$, the parameter $m_{\sigma}$ is not the mass of any propagating state~\cite{Achucarro:2010jv, Baumann:2011su, Shiu:2011qw, Achucarro:2012yr}.} If $m_{\sigma} \gg H$ then $\sigma$ may be integrated out, leading to an effective action for $\R$ that is characterized by a nontrivial sound speed $c_s$ determined by $\alpha$ and $m_{\sigma}$ as $c_s^{-2} = 1 + 2 \epsilon \alpha^2 / m_{\sigma}^2$~\cite{Tolley:2009fg, Cremonini:2010ua, Achucarro:2010da, Achucarro:2012sm, Pi:2012gf}. Another regime that has been studied vastly is the quasi-single field regime, in which $m_{\sigma} \sim H$, and $\epsilon \alpha^2 \ll H^2$~\cite{Chen:2009zp, Chen:2009we}.

We are interested in those cases in which the entropy field cannot be integrated out (i.e. it remains light) and its effect on the evolution of curvature perturbations is significant throughout the whole period of inflation. In principle, a very light perturbation $\sigma$, that remains coupled to $\R$ for a long time, could generate large isocurvature perturbations towards the end of inflation and have undesired consequences on the dynamics of the adiabatic mode $\R$. However, these expectations are premature. We will consider an ultra-light entropy perturbation $\sigma$ coupled to $\R$ in the way described in (\ref{new-action-quadratic0}) and show that, under certain circumstances, it provides interesting new physics compatible with current observations.

If we want to make $\sigma$ exactly massless, in the sense that the system is invariant under shifts of $\sigma$ by itself,  we are forced to take both $m_\sigma= 0$ and $\alpha=0$. However, in this case, the evolution of $\sigma$ will be completely decoupled from $\R$, leading to the production of isocurvature and curvature perturbations of comparable  sizes. Depending on the details of reheating, this could imply large levels of isocurvature perturbations in the CMB. However, there is an alternative formulation of (\ref{new-action-quadratic0}) which allows one to identify a sense in which $\sigma$ becomes effectively massless on very large scales. We can rearrange the terms in (\ref{new-action-quadratic0}) and rewrite it as
\be\label{new-action-quadratic-mu}
S=\int \! d^4x \, a^3  \left[\epsilon ( \dot \R -  \alpha \sigma)^2  - \frac{ \epsilon }{ a^2 } \, (\nabla \R)^2 + \frac{1}{2}\left(\dot\sigma^2-\frac{1}{a^{2}}(\nabla\sigma)^2\right)-\frac{1}{2} \mu^2\sigma^2\right] ,
\ee
where $\mu = \sqrt{m_{\sigma}^2 + 2 \epsilon \alpha^2}$ is the so called entropy mass of $\sigma$. The key point is that  when $\mu = 0$ this action has a St\"uckelberg-like symmetry that involves shifts of both fields $\sigma$ and $\dot \R$. This symmetry ensures that $\sigma$ behaves effectively as a massless field on super horizon scales. To see this, let us consider the long wavelength limit of~(\ref{new-action-quadratic-mu}), in which spatial gradients may be disregarded, and momentarily introduce the following non-local field redefinition:
\be\label{psi}
\dot \psi \equiv \dot \R-\alpha\sigma .
\ee
Then, on super-horizon scales,~(\ref{new-action-quadratic-mu}) becomes an action that describes the dynamics of two decoupled fields:
\be
S_{\rm sh}=\int d^4x a^3\left[\epsilon\dot\psi^2+{1\over 2}\dot\sigma^2 - \frac{1}{2} \mu^2 \sigma^2  \right] . \label{S_sh}
\ee
This action is explicitly invariant under shifts of $\psi$, and one may infer that the field $\sigma$ behaves as a field of mass $\mu$ on super-horizon scales. It follows that in the limit $\mu \to 0$ the field $\sigma$ becomes massless in the usual sense.\footnote{Note that the use of the field re-definition is only a trick to explicitly see the symmetries. One could instead work with $\R$ and $\sigma$ to obtain the same result.}  In terms of $\R$ and $\sigma$, the symmetry obtained in the zero entropy mass limit $\mu \to 0$ is equivalent to the following St\"uckelberg-like transformation:
\begin{eqnarray}
\dot \R \rightarrow \dot \R-\alpha \, \delta C_1 ,  \label{sym-R} \\
\sigma\rightarrow \sigma+\delta C_1 .\label{sym-sigma}
\end{eqnarray}
From now on we consider the case $\mu^2 = 0$ where $\sigma$ can be considered as an effectively massless field on super-horizon scales (see ref.~\cite{Renaux-Petel:2015mga} for a recent discussion on instabilities of the background when $\mu^2$ is negative).

Reintroducing  spatial gradients, although $\sigma$ can no longer be considered shift invariant by itself, we see that the combined symmetry \eqref{sym-R} and \eqref{sym-sigma} is a symmetry of the following full action
\be\label{new-action-quadratic}
S=\int d^4x a^3\left[\epsilon\left(\dot \R -\alpha\sigma\right)^2  - \frac{ \epsilon }{ a^2 } \, (\nabla \R)^2  -\frac{1}{2}\left(\dot\sigma^2  -  \frac{1}{a^{2}} (\nabla\sigma)^2\right)\right] .
\ee
We will now explore the situation in which $\sigma$ can be effectively considered as a massless perturbation at  sub-horizon scales. At very short wavelengths we can neglect the Hubble friction, in other words we consider modes of wave number $k^2 /a^2 \gg H$. In this case, we can consider the ``Minkowski" limit of the action \eqref{new-action-quadratic} (basically we fix $a\sim 1$). In this regime, at leading order in slow-roll, we see immediately that the coupling $\dot \R\sigma$ is invariant under $\sigma\rightarrow \sigma+$const (as it only generates a boundary, i.e. under this transformation the action is left invariant). We are still left with an explicit non-derivative term proportional to $\epsilon \alpha^2$ that would break this symmetry. Nevertheless, if $\epsilon \alpha^2 \ll H^2$, this term can be neglected on the same grounds as the Hubble friction is neglected.  In this case, $\sigma$ behaves as a massless field also on sub-horizon scales.

In fact, the system in this regime turns out to be secretly described by two exactly massless modes. In other words, under a specific wavelength-dependent rotation of $\R$ and $\sigma$, one can obtain two other shift invariant fields (for details see Appendix A of ref.~\cite{Castillo:2013sfa}) even for $\alpha\neq 0$. Therefore, combining with the super-horizon result, at least within our approximations, our system is precisely described by two massless modes. However, since we are interested in the physical $\R$ and $\sigma$, we will not further consider this direction.

\setcounter{equation}{0}
\section{Evolution of the adiabatic mode} \label{sec:evolution_of_R}

We now examine the long-wavelength evolution of the fields $\R$ and $\sigma$ as determined by the symmetries of the action (\ref{new-action-quadratic}). In Appendix~\ref{app:dynamics} we offer a more detailed analysis of the evolution of these fields paying special attention on their mode-decomposition. To start with, let us recall that in addition to the symmetry of eqs.~(\ref{sym-R}) and~(\ref{sym-sigma}), the system is invariant under shifts of $\R$:
\be
\R \to \R + \delta C_2 . \label{sym-c1}
\ee
On the other hand, the equations of motion for $\R$ and $\sigma$ derived from action~(\ref{new-action-quadratic}) are invariant under the set of transformations
\bea
 \dot \R &\to& \dot \R -\frac{\delta C_3}{\epsilon a^3} - \alpha \frac{\delta C_4}{a^3} \label{decay1}
 \ , \\
 \sigma &\to& \sigma -\frac{\delta C_4}{a^3}. \label{decay2}
\eea
These transformations may be used to identify the long wavelength solutions to the equation of motion for $\R$ and $\sigma$. In particular, the transformation involving $\delta C_3$ tells us that $\R$ has a solution that decays as $1 / \epsilon a^3$, whereas the transformation involving $\delta C_4$ informs us that $\sigma$ has a solution that decays as $1 / a^3$. Disregarding these decaying modes, the invariance of the system under (\ref{sym-c1}) implies that the adiabatic mode $\R$ has a constant solution $\R_0$ that becomes manifest on super horizon scales. By the same token, the invariance of the system under the transformation (\ref{sym-sigma}) involving $\delta C_1$ reveals that $\sigma$ has a constant solution $\sigma_0$. However, this transformation also forces $\R$ to have a growing solution on super-horizon scales that is dictated by $\sigma_0$. In other words, the long wavelength evolution of $\R$ is given by:
\be
\R \simeq  \sigma_0 \int^N_{N_0} \!\! dN \, \frac{\alpha}{H} + \R_0, \label{R-integral}
\ee 
where we have introduced $e$-folds $N$ (defined via $d N = H dt$). If $\alpha / H$ stays nearly constant, then the solution for $\R$ eventually becomes dominated by the particular growing solution proportional to $\sigma_0$, and the entire system is determined by a single degree of freedom on very long wavelengths. In this case, we obtain
\be
\R \simeq \frac{\alpha}{H}\sigma_0\Delta N+\R_0\ , \label{R-sigma_0}
\ee
where $\Delta N = N-N_0$. If $\R_0$ is the value of $\R$ at the time of horizon crossing (for a given mode with a wavelength that crosses the horizon at $N_0$), then $\Delta N$ is precisely the number of $e$-folds elapsed since horizon crossing.  Because towards the end of inflation $\Delta N$ is large (about $60$), the contribution to $\R$ proportional to $\sigma_0$ could dominate even if the ratio $\alpha/H$ is small. We further explore this situation in Section~\ref{sec:enhancing_power}.

\setcounter{equation}{0}
\section{A multi-field realization} \label{sec:simple-realization}

The action of eq.~(\ref{new-action-quadratic}) appears as a particular limit of multi-field models of inflation. Indeed, the most general action for a set of two scalar fields, with at most two space-time derivatives, is given by
\be
S = \frac{1}{2} \int \! d^4 x \sqrt{-g} \, R -  \int \! d^4 x \sqrt{-g} \left[ \frac{1}{2} \gamma_{a b} g^{\mu \nu} \nabla_{\mu} \phi^a \nabla_{\nu} \phi^b + V \right] , \label{action-two-field-basic}
\ee
where the first term corresponds to the usual Einstein-Hilbert action constructed out of the space-time metric $g_{\mu \nu}$ (with an inverse metric given by $g^{\mu \nu}$), $\gamma_{ab}$ is the $\sigma$-model metric characterizing a 2-dimensional target space spanned by the fields $\phi^1$ and $\phi^2$ (with an inverse metric given by $\gamma^{a b}$), and $V$ is the scalar field potential of the model. One may study the fluctuations of this system by first solving the background equations of motion, for fields $\phi_0^1$ and $\phi_0^2$, which are given by
\be
\frac{D}{dt} \dot \phi_0^a + 3 H \dot \phi_0^a + V^a = 0, \qquad 3 H^2 = \frac{1}{2} \dot \phi_0^2 + V, \label{equations-of-motion-scalars}
\ee
where $V^a \equiv \gamma^{a b} V_{b}$, and $D/dt$ represents a covariant derivative with respect to cosmic time $t$, whose action on an arbitrary vector $A^a$ is defined to be $D_t A^a \equiv \dot A^a + \Gamma^{a}_{b c} \dot \phi^b A^c$, where $\Gamma^{a}_{b c}$ are the usual Christoffel symbols computed out of $\gamma_{ab}$. To understand the influence of the background on the perturbations, it is useful to first define unit vectors tangent and normal to the trajectory~\cite{GrootNibbelink:2000vx, GrootNibbelink:2001qt}. In the case of $2$-field models, these may be respectively defined as~\cite{Cespedes:2012hu}:
\be
T^a \equiv \frac{\dot \phi^a }{\dot \phi_0} , \qquad N_a \equiv \sqrt {\det \gamma} \, \epsilon_{a b} T^b
\ee
where $\dot \phi_0 \equiv \sqrt{\gamma_{a b} \dot \phi_0^a \dot \phi_0^b}$, and $\epsilon_{ab}$ is the two-dimensional Levi-Civita symbol with $\epsilon_{11} = 1$. These unit vectors allow us to define a few relevant quantities that appear as couplings in the equations of motion for the perturbations $\R$ and $\sigma$ of the previous discussions. In the first place, it is crucial to define the turning rate $\Omega$ of the inflationary trajectory as:
\be
\Omega \equiv - N_a \frac{D T^a}{dt} .
\ee
Then, using the equations of motion~(\ref{equations-of-motion-scalars}), it is straightforward to find the following alternative relation for $\Omega$ in terms of first derivatives of the potential:
\be
\Omega = N^a V_a / \dot \phi_0.  \label{Omega-N-V}
\ee
Now, to study the dynamics of linear perturbations, it is useful to consider the Arnowitt-Deser-Misner (ADM) formalism~\cite{Arnowitt:1962hi} to write the full metric as
\be
ds^2 = - \N^2 dt^2 + a^2(t) e^{2 \R} \delta_{ij} (dx^i + \N^i)(dx^j + \N^j) , \label{metric}
\ee
where $\N$ and $\N^i$ are the lapse and shift functions, respectively, and $\R$ is the spatial curvature perturbation. We may write $\delta \phi^a (\x , t) = \phi^a(\x , t) - \phi^a_0 (t)$ and project this definition along the tangent and normal directions in such a way that:
\be
\delta \phi^a (\x , t) = T^a (t) \delta \phi_{||} (\x , t) + N^a (t) \sigma (\x , t) .
\ee
In this way $\delta \phi_{||} (\x , t)$ corresponds to the inflaton perturbations, parallel to the background trajectory, whereas $\sigma(\x ,t )$ corresponds to perturbations normal to the trajectory. In this formalism $\sigma$ represents the entropy perturbations (i.e. the non-adiabatic mode~\cite{Langlois:2008mn}). We may now adopt the co-moving gauge (sometimes also referred to as unitary gauge), whereby $\delta \phi_{||}(\x , t) = 0$. In this case the variable $\R$ introduced in the ADM splitting represents the adiabatic mode of perturbations~\cite{Langlois:2008mn}.
 
 After plugging the ADM metric back into the action~(\ref{action-two-field-basic}) one may solve the constraint equations to find $\N = 1 + \dot \R / H$ and $\N_i = \nabla_i ( \chi - \R / H)$, where $a^{-2} \nabla^2 \chi = \epsilon \dot \R  + \sqrt{2 \epsilon} \Omega \, \sigma $. These solutions then allow one to deduce that the action for the fluctuations $\R$ and $\sigma$, to quadratic order, is given by
\be
S = \int \! d^4 x \, a^3 \! \left[ \epsilon \dot \R^2 - \frac{\epsilon}{a^2} (\nabla \R)^2 + 2 \sqrt{2 \epsilon} \Omega  \dot \R \sigma + \frac{1}{2} \dot \sigma^2  - \frac{1}{a^2} (\nabla \sigma)^2 - \frac{1}{2} m_{\sigma}^2  \sigma^2 \right] ,\label{new-action-quadratic-0}
\ee
(exactly of the form \ref{new-action-quadratic0}) where the effective mass $m_{\sigma}$ is defined in terms of the projection of the second derivative of the potential along the normal direction $N^a$
\be
m_{\sigma}^2 \equiv N^{a} N^{b} ( V_{a b} - \Gamma_{ab}^c V_c ) + \epsilon H^2 \mathbb{R} - \Omega^2, \label{m_eff-def-1}
\ee
where $\mathbb{R}$ is the Ricci scalar characterizing the two-field manifold. To complete the identification it is enough to realize that the entropy mass $\mu$ and the parameter $\alpha$ introduced in Section~\ref{sec:coupling-R-sigma} correspond to the following combinations
\be
\mu^2 \equiv m_{\sigma}^2 + 4 \Omega^2,  \qquad  \alpha \equiv -  \frac{2 \Omega}{\sqrt{2 \epsilon}} . \label{m-def}
\ee
These identifications allow us to recover the action (\ref{new-action-quadratic-mu}). Moreover, in the limit $\mu \to 0$ we recover the desired action (\ref{new-action-quadratic}). More importantly, we now see what $\mu \to 0$ means in terms of the dynamics of the background. Concretely, this limit corresponds to the case in which $m_\sigma^2$ defined in (\ref{m_eff-def-1}) satisfies the condition:
\be
m_{\sigma}^2 = - 4 \Omega^2   .  \label{massless-condition}
\ee
As we shall see in Section~\ref{concrete-example}, this condition can arise in concrete multi-field setups without fine tuning.

\setcounter{equation}{0} 
\section{Enhancement of the power-spectrum} \label{sec:enhancing_power}

We now study the impact of the long wavelength evolution of $\R$, as examined in Section~\ref{sec:evolution_of_R}, on the spectra of the theory. In order to study the observables associated to this theory ---using tools from perturbation theory--- it is useful to define the following dimensionless quantity:
\be
\lambda  \equiv \sqrt{2 \epsilon} \alpha / H.
\ee
In terms of this coupling, action (\ref{new-action-quadratic}) now reads:
\be\label{action-final}
S = \int d^4 x a^3  \left[ \epsilon \left(\dot \R - \lambda \frac{H}{\sqrt{2 \epsilon}}  \sigma \right)^2 - \frac{\epsilon}{a^2} (\nabla \R)^2 + \frac{1}{2} \dot \sigma^2  - \frac{1}{a^2} (\nabla \sigma)^2 \right] .
\ee
Now, as discussed in Section~\ref{sec:evolution_of_R}, notice that if the number of $e$-folds elapsed since horizon crossing is large enough, then the growing mode of $\R$ dictated by $\sigma$ in eq.~(\ref{R-sigma_0}) will eventually dominate. If this indeed happens, then from eq.~(\ref{R-integral}) we see that $\R \simeq \sigma_0 \int^N_{N_*} dN \lambda / \sqrt{2 \epsilon}$, and the two point correlation function of $\R$ takes the form:
\be
\langle \R\R\rangle  =   \langle\sigma\sigma\rangle\ \left(  \int^N_{N_*} \!\!\!  dN'  \frac{\lambda}{\sqrt{2 \epsilon}} (N') \right)^2 \simeq \langle\sigma\sigma\rangle\   \frac{\lambda^2}{2 \epsilon} \Delta N^2 , \label{power-integral}
\ee
where we have assumed that $\lambda / \sqrt{2 \epsilon}$ is nearly constant throughout inflation.\footnote{In the formalism of \cite{Wands:2002bn}, where $\dot{{\cal R}}=\bar{\alpha}(t)H {\cal S}$ and $\dot{{\cal S}}=\bar{\beta}(t)H {\cal S}$, with ${\cal S}\equiv \sigma/(\sqrt{2\epsilon})$, this regime correspond to the limit $\bar{\alpha}/\sqrt{2\epsilon}={\rm const}$ and $\bar{\beta}=-\dot{\epsilon}/(2\epsilon H)$. In Section~\ref{concrete-example} we will introduce a concrete setup where these conditions are fully satisfied.} Here, $\Delta N = N - N_*$, where $N_*$ is the $e$-fold value at which the mode of interest crossed the horizon. A direct consequence of this relation is that the power spectrum of curvature perturbations $\R$ is determined by the power spectrum of the non-adiabatic perturbation $\sigma$ as:
\be
\P_{\R}=  \frac{\lambda^2}{2 \epsilon} \Delta N^2 \,  \P_{\sigma}  .
\ee
As discussed in Section~\ref{sec:coupling-R-sigma}, $\sigma$ approximately evolves as a massless field at both sub- and super-horizon scales. Therefore, its spectrum, within these approximations, must be given by the usual expression
\be
\P_{\sigma}\simeq \frac{H^2_*}{4\pi^2}\ ,
\ee 
where $H_*$ is the value of the Hubble scale at horizon crossing. This finally implies that the power spectrum for curvature perturbations is given by:
\be
\P_{\R}\simeq\frac{H^2_*}{8\pi^2\epsilon} \lambda^2\Delta N^2 . \label{spectrumR}
\ee 
Notice that while $H_*$ is evaluated at horizon crossing, the combination $\lambda^2 / \epsilon$ is evaluated towards the end of inflation. In addition, recall that $\Delta N = N - N_*$. These two considerations together imply that the spectral index does not depend on the running of $\lambda^2 / \epsilon$, and we obtain (see Appendix~\ref{app:observables} for details)
\bea\label{eq:ns}
n_s-1 = - 2 \left( \epsilon  + \frac{1}{\Delta N}  \right) . 
\eea
We would like to stress here that, given the way we have computed the spectrum for $\sigma$, it does not depend on $\lambda$ (for the exact calculation see Appendix~\ref{sec:pheno}). Of course that is not strictly correct as there will be a weak $\lambda$ dependence in the transition between sub to super horizon scales, that dies off in the super horizon limit, as well as a $\lambda$ dependence on the next to leading order term in the slow roll parameters.

If $\lambda = 0$, then the constant $\R_0$ in eq.~(\ref{R-sigma_0}) is responsible for determining the value of the power spectrum at the end of inflation, and would obtain the conventional result
\be
\P_\R  \Big|_{\lambda = 0} = \frac{H^2_*}{8 \pi^2 \epsilon_*} ,
\ee
where $\epsilon_*$ is the value $\epsilon$ at horizon crossing. Comparing this result with the previous expression (\ref{spectrumR}) reveals that the coupling $\lambda$ plays a role only if 
\be
\lambda^2 \Delta N^2 \gg 1 . 
\ee
We would like to stress here a couple of interesting features. First of all, the power-spectrum $\P_\sigma$ of the primordial non-adiabatic perturbations is not enhanced by cumulative effects (i.e. by the $\lambda^2 \Delta N^2$ factor). Therefore, regardless of the reheating mechanism connecting primordial non-adiabatic perturbations to the non-adiabatic perturbations in the CMB, those will be suppressed by at least a factor $\lambda^2 \Delta N^2$ with respect to the adiabatic perturbations. In second place, because the tensor spectrum is not enhanced, the tensor to scalar ratio $r$ will be suppressed too with respect to the standard single field scenario. In other words, as we can already guess (see Appendix~\ref{app:observables} for details):
\be
r\simeq \frac{16\epsilon}{\lambda^2 \Delta N^2}\ll r_{\rm single}=16\epsilon\ ,
\ee
where $r_{\rm single}$ is the equivalent tensor-to-scalar ratio in the single field scenario.

\setcounter{equation}{0}
\section{A concrete example} \label{concrete-example}

In this section we offer a concrete 2-field model where an ultra-light field coupled to the inflaton emerges with all the properties discussed in the previous sections. Specifically, this model is characterized by a background trajectory that satisfies $m_{\sigma}^2  = - 4 \Omega^2$, which, as we say in Section~\ref{sec:simple-realization}, is the required condition to ensure the existence of the ultra-light field. The model to consider has the following action:
\be
S = \frac{1}{2} \int d^4 x R -  \int d^4 x \left[ \frac{1}{2} e^{2 \Y / R_0} (\nabla \X)^2 + \frac{1}{2} (\nabla \Y)^2  + V (\X ) \right] . \label{action-X-Y}
\ee
Notice that this model has a field metric describing a 2-dimensional hyperbolic manifold of curvature $-2/R_0^2$, spanned by the fields $\X$ and $\Y$. Now, it may be noticed that when $V$ is independent of both $\X$ and $\Y$, the above action has a symmetry given by $\Y \to \Y' = \Y + C$ and $\X  \to  \X' =  e^{-C / R_0} \X$. If the potential is flat enough, as required by inflation, this symmetry is only weakly broken and will lead to the approximate symmetry of eqs.~\eqref{sym-R} and~\eqref{sym-sigma}. 

We are now ready to analyze the system as in Section~\ref{sec:simple-realization}. The metric in field space is given by:
\be
\gamma_{ab} = \left(\begin{array}{cc}  e^{2 \Y / R_0} & 0 \\0 & 1 \end{array}\right) .
\ee
From it, the non-vanishing Christoffel symbols are then found to be $\Gamma^{\X}_{\X \Y} = \Gamma^{\X}_{\Y \X} = 1 / R_0$, and $\Gamma^{\Y}_{\X \X} = - e^{2 \Y / R_0} / R_0$. As a consequence, the Ricci scalar is found to be negative and constant $\mathbb{R} = - 2/R_0^2$. The background equations of motion are given by:
\bea
\ddot \X + 3 H \dot \X + \frac{2}{R_0} \dot \X \dot \Y + e^{- 2 \Y / R_0}  V_{\X} = 0  , \\
\ddot \Y + 3H \dot \Y -  \frac{1}{R_0} e^{2 \Y / R_0} \dot \X^2 = 0 .
\eea
Models of this type, with a non-trivial kinetic term, have been studied elsewhere~\cite{GarciaBellido:1995fz, DiMarco:2002eb, DiMarco:2005nq, Lalak:2007vi, Cremonini:2010sv, Cremonini:2010ua, vandeBruck:2014ata}, and are well motivated from supergravity and string theory EFT's. Some of the results presented in this section overlap with those found in ref.~\cite{Cremonini:2010sv}, where the generation of iso-curvature perturbations due to a non-trivial kinetic term were studied. Given that the potential is independent of $\Y$, then the system is found to have a slow-roll regime of the form:
\be
 3 H \dot \X \simeq - e^{- 2 \Y / R_0}  V_{\X}  , \qquad
 3H \dot \Y \simeq \frac{1}{R_0} e^{2 \Y / R_0} \dot \X^2  , \qquad 
 3 H^2 \simeq V . \label{eom-approx}
\ee
This regime is characterized by the hierarchy $ \dot \Y^2 \ll e^{ 2 \Y / R_0}  \dot \X^2$. We deduce an expression for $\Y$ in terms of the potential $V(\X)$ given by
\be
 \Y \simeq  \Y_0 - \frac{1}{3 R_0}  \ln \left( \frac{V(\X)}{V(\X_0)} \right)  , \label{Y-VX}
\ee
where $\X_0$ and $\Y_0$ are the values of the fields at a given reference time $t_0$. In addition, one can obtain the following equation of motion for $\X$
\be
 \int^{\X}_{\X_0} \!\! d \X \left[  u (\X) \right]^{1 - \frac{2}{3 R_0^2} } \frac{ 1 }{d u / d \X} = -  e^{-2 \Y_0 / R_0}  ( N - N_0) , \label{X-V}
\ee
where $N - N_0$ is the number of $e$-folds that takes to go from $\X_0$ to $\X$, and where we have defined:
\be
u (\X) \equiv \frac{V(\X)}{V(\X_0)} .
\ee
This result can be used to obtain $\X$ as a function of $e$-folds $N$. Now, we may use the previous equations to find useful expressions for the tangent and normal vectors $T^a$ and $N^a$ valid in the slow-roll approximation. These are found to be given by
\be
T^a \simeq e^{- \Y / R_0} \left(1 , -  \frac{V_{\X}}{9 H^2 R_0 }    \right) , \qquad  N^a \simeq    -\left(  \frac{ e^{- 2 \Y / R_0}  V_{\X}}{9 H^2 R_0} ,   1 \right) . \label{T-and-N}
\ee
Now, using eq.~(\ref{Omega-N-V}) we find the following relation for $\Omega$:
\be
\Omega^2 \simeq  \frac{ e^{- 2 \Y / R_0}  V_{\X}^2}{9 H^2 R_0^2} .
\ee
From this result, it is easy to find that $\epsilon$ and $\Omega$ respect the following relation:
\be
\epsilon =   \frac{\Omega^2 R_0^2}{2 H^2} .
\ee
Then, putting all of these results back into eq.~(\ref{m_eff-def-1}), we finally obtain $m_{\sigma}^2 = - 4 \Omega^2$. This result shows that the system has two light fields, as promised. 

\subsection{Monomial potentials $V(\X) = \alpha (\X / \X_0 )^{n} $}

To have a more concrete picture of the background dynamics in the present class of models, let us consider the case of monomial potentials of the form
\be
V(\X) = \alpha (\X / \X_0 )^{n} , \label{V-mono}
\ee
where, as in the previous section, $\X_0$ is the value of the field $\X$ at a given reference time $t_0$. Unless the power $n$ is an even number, the potential of eq.~(\ref{V-mono}) is not suitable to properly finish inflation. To address the end of inflation one may consider the alternative potential\footnote{We thank Pablo Gonzalez and Nelson Videla for bringing up this extension to us.} $V(\X) = \alpha \left[  (b + (\X / \X_0)^2 )^{n/2} - b^{n/2} \right]$ with $b \ll 1$. This potential approximates to $\alpha (\X / \X_0 )^{n}$ for values of $\X$ around $\X_0$, and behaves well around the minimum $\X = 0$. For simplicity, we continue working with eq.~(\ref{V-mono}). Then, eq.~(\ref{Y-VX}) becomes:
\be
 \Y = \Y_0 -\frac{n}{3 R_0} \ln (\X / \X_0 ) .
\ee
On the other hand, integrating eq.~(\ref{X-V}) we obtain $\X$ as a function of $N$ via the following relation
\be
 \frac{ \X_0^2}{2 - \frac{2 n}{3 R_0^2}} \left[ \left( \frac{\X}{\X_0} \right)^{2 - \frac{2 n}{3 R_0^2}} - 1  \right]\simeq - n  e^{- 2 \Y_0 / R_0} (N - N_0) ,
\ee
where $N_0$ is the number of $e$-folds at the reference time $t_0$. Now, notice that if $R_0^2 >  n / 3$ then the field $\X$ will be able to reach $\X = 0$ after a finite number of $e$-folds $\Delta N = N_{\rm end} - N_0$. This gives us back a relation determining the amount of $e$-folds $\Delta N$ that the system takes to reach the end of inflation (at $\X = 0$) from the configuration $(\X_0 , \Y_0)$:
\be
n \left( 2 - \frac{2 n}{3 R_0^2} \right) \Delta N = \X_0^2 e^{2 \Y_0 / R_0}  . \label{X0Y0-N}
\ee
Notice that there is a degeneracy on the values of the parameters $\X_0$ and $\Y_0$ allowing for a certain fixed amount of $e$-folds, determined by the equation $\X_0^2 e^{2 \Y_0 / R_0} =$constant. To continue, eq.~(\ref{X0Y0-N}) allows us to determine the time dependence of relevant background quantities, such as $\lambda$ and $\epsilon$. Concretely, we find that $\lambda$ and $\epsilon$ have the following dependence on $\Delta N$:
\be
\lambda = \sqrt{  \frac{ 6 n }{   \left( 3 R_0^2 - n \right) \Delta N}  } , \qquad \epsilon = \frac{  n }{4 \left( 1 - \frac{ n}{3 R_0^2} \right) \Delta N} .
\ee
It is important to notice that the combination $\lambda / \sqrt{2 \epsilon}$ is a constant (independent of the number of $e$-folds) given by:
\be
\frac{\lambda}{\sqrt{2 \epsilon}} =  \frac{2}{ R_0} .
\ee
Recall that $\lambda / \sqrt{2 \epsilon} = \alpha /H$ determines the coupling appearing in action~(\ref{new-action-quadratic}). As a result, in this class of models, the observables discussed in the Appendix~\ref{sec:pheno} are accurately given by the various expressions thereby deduced. In particular, in this model, one finds that in the spectral index ($n_s$) the running of $\lambda$ cancels with $\eta$, and therefore, from (\ref{eq:ns}),
\be
n_s - 1 = - \frac{1}{\Delta N} \left[2 +  \frac{  n }{2 \left( 1 - \frac{ n}{3 R_0^2} \right) }  \right] . \label{ns_large_coup}
\ee
On the other hand, the tensor to scalar ratio $r$ is found to be
\be
r = \frac{2 R_0^2}{\Delta N^2} ,\label{r_large_coup}
\ee
which is independent of the power $n$. Now, if we wish to have a spectral index given by $n_s \simeq 0.96$ and $\Delta N = 60$ $e$-folds of inflation, we see that the following relation between $R_0$ and $n$ has to be fulfilled:
\be
R_0^2 =  \frac{4 n}{12 - 15 n} .
\ee
This result tells us that $n < 4/5$, which implies that the potential $V$ must be concave in order to have a spectral index compatible with current observations. For instance, in the particular case $n = 1/2$ we require $R_0 = 2/3$. This in turn implies that $\lambda_{*} = \sqrt{ 3/ 50  }   \simeq 0.245$. With this value, we see that $\lambda_*^2\Delta N^2\sim 2\times 10^2$ which provides a huge enhancement to the power spectrum. Finally, the value of $\P_\R$ is determined by $\alpha$, and is fixed by the COBE normalization of the modes.

The expressions for $n_s$ and $r$ given above are valid in the case $\lambda^2 \Delta N^2 \gg 1$, in which the curvature perturbations at the end of inflation are dominated by the mode that grows after horizon crossing. If however $\lambda^2 \Delta N^2 \ll 1$, the curvature and isocurvature fields are effectively decoupled, and the statistical properties of the curvature perturbations are determined solely at horizon crossing. In terms of the $(n_s,r)$ plane, the latter case reduces to a single-field model of inflation (while it doesn't from the point of view of the curvature to isocurvature fluctuations). For monomial potentials $V=\alpha\phi^n$, the spectral index and tensor to scalar ratio are given by
\bea
n_s-1&=& \frac{2(2+n)}{n+4N_{\text{tot}}}\label{eq:ns_single_field} \ ,\\
r&=&\frac{16n}{n+4N_{\text{tot}}} \label{eq:r_single_field} \ .
\eea
As a function of $R_0$, the predictions in the $(n_s,r)$ plane will then interpolate between the `large coupling' regime given in eqs.~(\ref{ns_large_coup})-(\ref{r_large_coup}) and the single field predictions given in eqs.~(\ref{eq:ns_single_field})-(\ref{eq:r_single_field}).

\setcounter{equation}{0}
\section{Conclusions} \label{sec:conclusions}

Muti-field models of inflation with several light fields such as pseudo-Nambu-Goldstone bosons or moduli are fairly common in string theory and beyond the standard model cosmological scenarios. It is important to understand to what extent they are compatible with the observations.

In this paper we have considered multi-field theories of inflation in which the inflaton fluctuations interact with light fields in such a way that (1) isocurvature modes are generated during inflation, and (2) curvature perturbations $\R$ experience a large super-horizon growth sourced by the light fields. Contrary to expectations, there is a very interesting regime in which the prolonged combination of these two properties can lead to an effective suppression of isocurvature perturbations at the end of inflation, while still retaining a nearly scale invariant power spectrum. As a byproduct, we show that these models also provide a natural mechanism to suppress the tensor to scalar ratio compared to the single field realization with the same spectral index, and slow-roll parameters. Previous works~\cite{Kobayashi:2010fm, Cremonini:2010sv, Cremonini:2010ua} have analyzed the effects of light non-adiabatic fields on the evolution of curvature perturbations and found similar results. In this work, however, we have attempted to give various steps forward in understanding this class of models in a more systematic and generic way. In particular, (a) we have identified  a novel, non-standard  shift symmetry relating the entropy and curvature perturbations,  that emerges in the ultra-light limit $\mu \to 0$. (b) We presented a concrete model (discussed in Section~\ref{concrete-example}) where this symmetry is realised. We have  discussed the role of the underlying symmetries of the scalar field manifold  in ensuring that both the shift symmetry and the sourcing of  curvature by entropy perturbations persist during the entire history of inflation.  (d) We have analytically solved the system of perturbations in the regime in which this shift symmetry is valid (see Appendix~\ref{app:quantization}).
 
The key mechanism discussed here is the following: Since the entropy mode is approximately massless, it freezes on super-horizon scales acting as a constant source for the curvature perturbation. If inflation last enough $e$-folds, the adiabatic mode would then linearly grow in time. Remarkably, this implies that at late times,  both the curvature as well as the isocurvature perturbations are effectively dominated by a single degree of freedom. It is interesting to note that the effective field theory description of this regime is expected to be quite different from~\cite{Senatore:2010wk} and/ or the curvaton scenario, and may appear in the context of scenarios with a large number of fields, where a given combination of them could couple to the adiabatic curvature perturbation in the form of a light field~\cite{Dias:2016slx}.

If the entropy mode remains massless for a sufficiently large number of $e$-folds $\Delta N\gg 1$,  and the coupling  is perturbatively small, the power spectrum of isocurvature perturbations evolves in the standard way while curvature perturbations $\P_\R$ are enhanced by a factor proportional to $\Delta N$. Effectively the primordial isocurvature modes are suppressed as
\be
\P_\R / \P_\S \simeq (\P_\R / \P_\C)^2 \propto \Delta N^2  \ee
where $\P_\S$ is the power spectrum of the normalized entropy perturbation~\cite{Amendola:2001ni} related to $\sigma$ by $\S\equiv \sigma / \sqrt{2\epsilon}$ and $\P_\C$ the cross-correlation spectrum $\langle \R \S \rangle$, calculated in Appendix~\ref{sec:pheno}, which is also suppressed.

These features seem to be tightly connected to an emergent symmetry: Besides the usual shift symmetry that is associated with the masslessness of $\R$ (the adiabatic mode), we have identified in the quadratic action a second one involving both the curvature and the ultra-light (entropy) mode $\sigma$. Specifically, to reach the regime we are interested in, the time derivative $\dot \R$ must appear in the following combination 
\be
\dot \R - \lambda\frac{H}{\sqrt{2 \epsilon}}\sigma   ,\label{DR}
\ee
where in the case of multi-field inflation the dimensionless coupling $\lambda$ is proportional to the rate of bend $\Omega$ of the inflationary trajectory ($\lambda = - \Omega /H$). The combination \eqref{DR} is invariant under the St\"uckelberg-like transformation
\be\label{stu2}
\R\rightarrow \dot \R - \lambda\frac{H}{\sqrt{2 \epsilon}}\delta C_2 ,Ê \qquad \sigma\rightarrow\sigma+\delta C_2\ .
\ee
There is some evidence, in concrete multi-field examples,  that this structure is inherited from a symmetry of the parent theory. It is tempting to speculate that the combination of eq.~(\ref{DR}) does not only happen at quadratic order but will also appear at higher orders, with interesting consequences for non-Gaussianity. We leave this for future work.

\section*{Acknowledgments} 

We are grateful to Pablo Gonzalez, Sander Mooij, Jorge Nore\~na, Grigoris Panotopoulos, Spyros Sypsas, Krzysztof Turzynski, Nelson Videla, David Wands and Yvette Welling for useful comments and discussions. This work was partially supported by a Fondecyt project 1130777 (GAP) and by a grant from the Simons Foundation (AA). CG is supported by the Ramon y Cajal program and partially supported by the Unidad de Excelencia María de Maeztu Grant No. MDM-2014-0369 and FPA2013-46570-C2-2-P grant. AA and VA are partially supported by the Netherlands Organization for Scientific Research (NWO) and the Dutch Ministry of Education, Culture and Science (OCW). AA acknowledges support from the Basque Government (IT-559-10),  and Spanish Ministry MINECO  (FPA2015-64041-C2-1P).

\appendix

\numberwithin{equation}{section}

\section{Dynamics} \label{app:dynamics}

\begin{figure}[t!]
\begin{center}
\includegraphics[scale=0.36, bb=630 00 630 415]{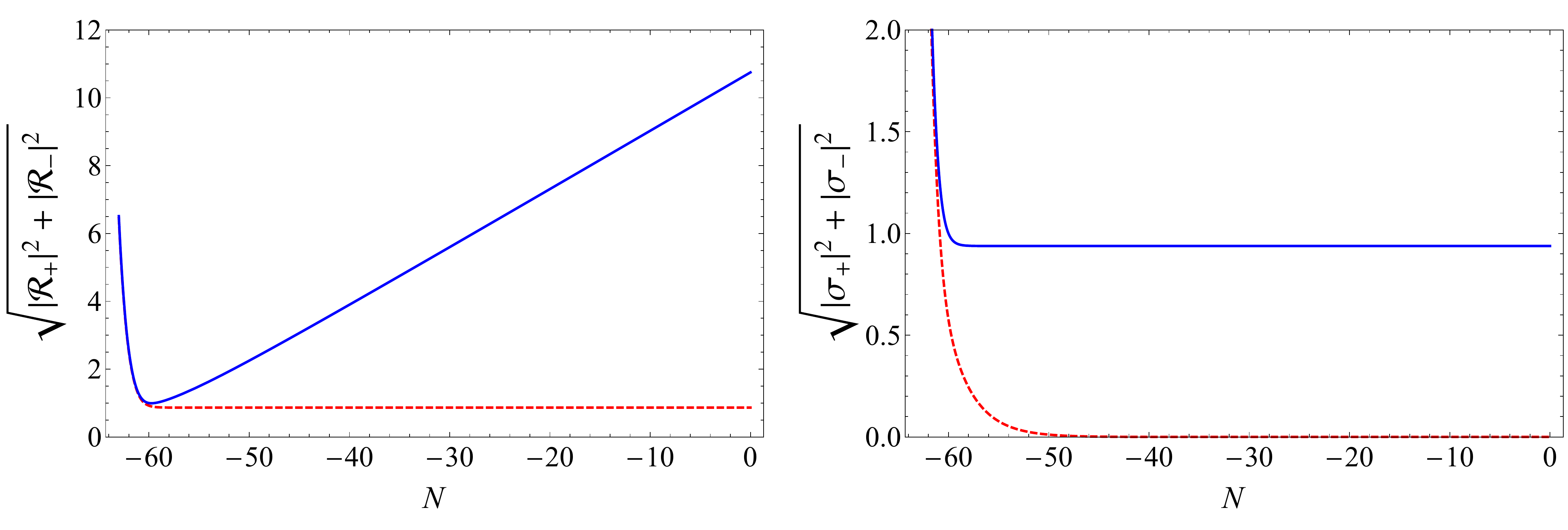}
\caption{The figure shows the evolution of the amplitude of the fluctuations around the time of horizon crossing (at around $N = -60$). The left panel shows the amplitude of $\R$, whereas the right panel shows the amplitude of $\sigma$. The red dashed curves correspond to the case in which there is no coupling between $\R$ and $\sigma$ (that is $\lambda = 0$), and $\sigma$ has a nonzero entropy mass $\mu$. It may be seen that $\R$ freezes whereas $\sigma$ decays quickly once they cross the horizon. The blue solid lines show the case in which the two fields remain coupled, with $\lambda = 0.2$, and $\sigma$ has zero entropy mass. In this case, $\R$ grows outside the horizon, and $\sigma$ freezes.}
\label{fig:R-Sigma}
\end{center}
\end{figure}

In this appendix we examine the dynamics of the fluctuations by studying the system of coupled equations of motion that emerge from the action~(\ref{action-final}). Throughout this analysis we will assume that every background quantity evolves slowly, including the coupling $\lambda$, which is assumed to satisfy:
\be
\big|  \dot \lambda / H \lambda \big| \ll 1 .
\ee
This assumption will allow us to take $\lambda$ as a constant, for all practical purposes, in computations that require us to take into account the evolution of fluctuations. To start with, let us recall that during inflation, perturbations are dominated by quantum fluctuations that are stretched from sub-horizon to super-horizon scales. These fluctuations respect the usual commutation relations of quantum mechanics, and satisfy Bunch-Davies initial conditions imposed on sub-horizon scales. To study the linear evolution of the coupled system of fluctuations, let us write both $\R$ and $\sigma$ in terms of Fourier modes as
\bea
\R (\x , t) &=& \!\! \int \frac{d^3k}{(2 \pi)^{3}} \hat \R (\k , t) e^{i {\bf k \cdot x}}  ,  \quad  \hat \R (\k , t) = \sum_{\alpha}   \left[ a_{\alpha}({\bf k}) \R_{\alpha} (k,t) + a_{\alpha}^{\dag}(-{\bf k}) \R^*_{\alpha}(k,t) \right] , \quad \label{R-k} \\
\sigma (\x , t) &=& \!\! \int \frac{d^3k}{(2 \pi)^{3}} \hat \sigma (\k , t) e^{i {\bf k \cdot x}}  ,  \quad  \hat \sigma (\k , t) = \sum_{\alpha}   \left[ a_{\alpha}({\bf k}) \sigma_{\alpha} (k,t) + a_{\alpha}^{\dag}(-{\bf k}) \sigma^*_{\alpha}(k,t) \right] , \quad  \label{sigma-k}
\eea
where $\alpha = +,-$ labels the two scalar modes, and $a_{\alpha}({\bf k})$ and $a_{\alpha}^{\dag}({\bf k})$ are creation and annihilation operators satisfying the following commutation relations:
\be
\big[ a_{\alpha}({\bf k}) , a_{\beta}^{\dag}({\bf k}) \big] = (2 \pi)^3 \delta_{\alpha \beta} \delta^{(3)} ({\bf k} - {\bf k}') . \label{comm-a}
\ee
The vacuum state of the system $| 0 \rangle$ is such that $a_{\pm} | 0 \rangle = 0$, and mode functions $\R_{\alpha} (k,t)$ and $\sigma_{\alpha} (k,t)$, which describe the evolution of fluctuations in momentum space, have amplitudes such that the quantum commutation relations between the fields, and their canonical momenta, are satisfied (see the Appendix~\ref{app:quantization} for more details). At linear order, each mode $\alpha$ evolves independently, with equations of motion given by
\bea
\frac{d}{dt} \left(\dot \R_{\alpha} - \lambda \frac{H}{\sqrt{2 \epsilon}} \sigma_{\alpha} \right) + (3 + \eta) H \left( \dot \R_{\alpha} - \lambda \frac{H}{\sqrt{2 \epsilon}} \sigma_{\alpha} \right)  + \frac{k^2}{a^2}Ê\R_{\alpha} &=& 0, \label{eom-R} \\
\ddot \sigma_{\alpha} + 3 H \dot \sigma_{\alpha} + \frac{k^2}{a^2} \sigma_{\alpha} + \sqrt{2 \epsilon} \lambda H \left(  \dot \R_{\alpha} - \lambda \frac{H}{\sqrt{2 \epsilon}}  \sigma_{\alpha} \right) &=& 0  , \label{eom-sigma}
\eea
where we have omitted the labels $k$ and $t$ for convenience. It is important to emphasize that both $\R$ and $\sigma$ are necessarily linear combinations of the modes labeled with $\alpha = +,-$ only as a result of the non-vanishing coupling $\lambda$. To be more concrete, if $\lambda = 0$, we the system of equations decouple, and we would be able to identify each mode with each field, be requiring, for instance, that $\R_{+} = \sigma_{-} = 0$ at all times. As soon as $\lambda \neq 0$, both $\R$ and $\sigma$ become a mixture of the two modes $+$ and $-$. As we shall see in the next section, well inside the horizon, where $k^2 / a^2 \ll H^2$, the role of the coupling $\lambda$ becomes negligible unless $\lambda \gg 1$, and we obtain a set of equations of motion describing fluctuations in a Minkowski space-time (allowing us to impose Bunch-Davies conditions). As soon as the modes approach the horizon, $\lambda$ will start to play a more important role, affecting the long wavelength behavior of the modes. We will see that the equations of motion imply that $\sigma$ freezes and that $\R$ grows linearly with respect to $e$-folds.  Figure~\ref{fig:R-Sigma} shows the evolution of the amplitudes $\sqrt{| \R_{-} (k,t) |^2 +  | \R_{+} (k,t) |^2}$ and $\sqrt{| \sigma_{-} (k,t) |^2 +  | \sigma_{+} (k,t) |^2}$ (normalized to their values at horizon crossing) as a function of $e$-folds, for modes with Bunch-Davies initial conditions and $\lambda = 0.2$. 

\subsection{Short wavelength behavior} \label{app:short}

The short wavelength limit of this system is characterized by fluctuations $\R_{\alpha}^{(s)}$ and $\sigma_{\alpha}^{(s)}$ such that their variations satisfy the following hierarchy:
\be
\dot \R_\alpha^{(s)} \gg H \R_\alpha^{(s)} , \qquad \dot \sigma_\alpha^{(s)} \gg H \sigma_\alpha^{(s)} .
\ee
To study this limit, it is useful to adopt conformal time $\tau$, which is determined by the relation $d\tau = dt / a$. In addition, we may define two new fields $u$ and $v$ out of $\R$ and $\sigma$, given by: 
\be
u  = a \sqrt{2 \epsilon} \R , \qquad v = a \sigma .
\ee
Then, the equations of motion~(\ref{eom-R}) and~(\ref{eom-sigma}) reduce to:
\bea
u_\alpha '' + \frac{\lambda}{\tau} v_\alpha'  + k^2Êu_\alpha - \frac{2}{\tau^2} u_\alpha - \lambda \frac{2 }{\tau^2} v_\alpha &=& 0, \\
v_\alpha ''  + k^2 v_\alpha - \frac{2}{\tau^2} v_\alpha  - \frac{\lambda}{\tau} \left(  u_\alpha' +  \frac{1}{\tau} u_\alpha + \frac{\lambda}{\tau} v_\alpha \right) &=& 0  .
\eea
To deal with this system we may perform the following $k$-dependent reparametrization of the fields, in order to obtain a new pair of fields $\Phi = (\bar u , \bar v)$ given by
\be
\Phi  \equiv \left(\begin{array}{c} \bar u_\alpha \\ \bar v_\alpha \end{array}\right) = \left(\begin{array}{cc} \cos \theta_k(\tau)& \sin \theta_k(\tau) \\ -\sin \theta_k(\tau) & \cos \theta_k(\tau)\end{array}\right) \left(\begin{array}{c}  u_\alpha \\  v_\alpha \end{array}\right) , \qquad \theta_k(\tau) = \frac{\lambda}{2} \ln (- k \tau) . \label{Phi}
\ee
This transformation corresponds to a rotation by an angle $\theta_k (\theta)$.  With this transformation, the equations of motion become
\bea
\Phi '' + k^2 \Phi + M^2 (\tau) \Phi = 0 ,  \label{Phi-eq}
\eea
where the mass matrix $M^2 (\tau) $ is found to be given by:
\be
M^2 (\tau) = - \frac{1}{\tau^2} \left(\begin{array}{cc} 2 + \frac{3}{4}\lambda+ \frac{3}{2} \lambda \sin 2 \theta  -  \lambda^2 \cos 2 \theta  &  \lambda \cos 2 \theta +  \lambda^2 \sin 2 \theta   \\  \lambda \cos 2 \theta  +  \lambda^2 \sin 2 \theta  & 2 + \frac{3}{4}\lambda - \frac{3}{2} \lambda \sin 2 \theta + \lambda^2 \cos 2 \theta  \end{array}\right) .
\ee
Equation~(\ref{Phi-eq}) allows us to be more precise about what we mean by the short wavelength regime. Indeed, comparing the entries of the mass matrix with $k^2$, we see that the short wavelength limit may be formally defined as the regime characterized by $k |\tau| \gg \max (1 , \lambda^2 )$. In this limit, the elements of the mass matrix become negligible compared to $k^2$, and we recover a set of fluctuations in a Minkowski space-time. This allows us to impose Banch-Davies conditions, which read:
\be
\bar u_- = \frac{1}{\sqrt{2 k}} e^{- i k \tau} , \qquad \bar u_+ = 0 , \qquad \bar v_+ = \frac{1}{\sqrt{2 k}} e^{- i k \tau} , \qquad \bar v_- = 0 .
\ee
Equation~(\ref{Phi-eq}) tells us that the effect of $\lambda$ will become relevant, as the modes approach the horizon. If $|\lambda| \gg 1$, then the coupling starts affecting the evolution of the system at a comoving scale determined by $k \sim  \lambda^2 a H$. Otherwise, if $|\lambda|$ is of order $1$ or smaller, then the coupling will become operative at a comoving scale $k \sim  |\lambda| a H$. In particular, if $|\lambda| \ll 1$, the effect of the coupling will be negligible during the entire short wavelength regime, and will only affect the long wavelength regime of the system. In the following section, we discuss this regime.

\subsection{Long wavelength behavior}

Let us now examine the dynamics of the system in the formal limit $k/a \to 0$, which corresponds to the long wavelength regime, after the modes have crossed the horizon. In the long wavelength limit, the system is characterised by fluctuations $\R_{\alpha}^{(\ell)}$ and $\sigma_{\alpha}^{(\ell)}$ satisfying the following equations of motion:
\bea
\frac{d}{dt} \left(\dot \R_{\alpha}^{(\ell)}  - \lambda \frac{H}{\sqrt{2 \epsilon}} \sigma_{\alpha}^{(\ell)}  \right) + (3 + \eta) H \left( \dot \R_{\alpha}^{(\ell)}  - \lambda \frac{H}{\sqrt{2 \epsilon}} \sigma_{\alpha}^{(\ell)}  \right)   &=& 0, \\
\ddot \sigma_{\alpha}^{(\ell)}  + 3 H \dot \sigma_{\alpha}^{(\ell)}   + \sqrt{2 \epsilon} \lambda H \left(  \dot \R_{\alpha}^{(\ell)}  - \lambda \frac{H}{\sqrt{2 \epsilon}}  \sigma_{\alpha}^{(\ell)}  \right) &=& 0  .
\eea
Now, the first of these two equation may be integrated once, to give:
\be
 \dot \R_{\alpha}^{(\ell)}  - \lambda \frac{H}{\sqrt{2 \epsilon}} \sigma_{\alpha}^{(\ell)}   = \frac{C_3^{(\ell)} (\alpha)}{a^3 \epsilon} , \label{R-C3}
\ee
where $C_3^{(\ell)} (\alpha)$ is an integration constant that may depend on the labels $\alpha$ and $\ell$. This result shows that the combination $ \dot \R_{\alpha}^{(\ell)}  - \frac{\lambda H}{\sqrt{2 \epsilon}} \sigma_{\alpha}^{(\ell)} $ dilutes exponentially with time as the universe inflates. Plugging this result back into the second equation of motion, we find that 
\be
\ddot \sigma_{\alpha}^{(\ell)}  + 3 H \dot \sigma_{\alpha}^{(\ell)}  = - \sqrt{\frac{2}{\epsilon}} \lambda H \frac{C_{3} ^{(\ell)}(\alpha)}{a^3}   .
\ee 
Since we are interested in the limit where $k / a \to 0$, we may neglect the term at the right hand side. The result is the equation of motion for a massless field $\ddot \sigma_{\alpha}^{(\ell)}  + 3 H \dot \sigma_{\alpha}^{(\ell)} = 0$. Then, integrating this equation twice, we find that $\sigma$ has a long wavelength evolution given by
\be
\sigma_{\alpha}^{(\ell)}   = C_2^{(\ell)}(\alpha) + C_4^{(\ell)}(\alpha) / a^3 ,
\ee
where $C_2^{(\ell)}(\alpha)$ and $C_4^{(\ell)}(\alpha)$ are integration constants. Finally, putting this result back into eq.~(\ref{R-C3}), and neglecting the term proportional to $C_4$, which also dilutes as space expands, we obtain:
\bea
\R_\alpha^{(\ell)}   &=& C_1^{(\ell)}(\alpha) +  C_2^{(\ell)}(\alpha)  \int^t \!\!\! dt' \frac{\lambda H}{\sqrt{2 \epsilon}} (t') , \label{R-super-h} \\
\sigma_{\alpha}^{(\ell)}  &=& C_2^{(\ell)}(\alpha) .  \label{sigma-super-h} 
\eea
The values of the integration constants are determined by the Bunch-Davies initial conditions for modes at sub-horizon scales discussed in Section~\ref{app:short}. Given that we are assuming that $\lambda$, $H$ and $\epsilon$ evolve slowly, we find that in these types of systems $\sigma_{\alpha}$ freezes at horizon crossing, and $\R_\alpha$ continues evolving linearly with respect to time. These results also show that the long wavelength behavior of $\R$ and $\sigma$ are consistent with the symmetries of the quadratic action under the transformation of eqs.~(\ref{sym-R}) and (\ref{sym-sigma}). Eqs.~(\ref{R-super-h}) and~(\ref{sigma-super-h}) are valid for arbitrary values of $\lambda$ (provide we can neglect higher-order contributions). In the next section, we will derive the coefficients $C_1$ and $C_2$ in the case $\lambda \ll 1$ and constant. As we explain below, these results can be used as boundary conditions for computing the coefficients $C_1$ and $C_2$ in the more general case in which $\lambda$ is allowed to vary in time and reach values $\mathcal{O}(1)$.

\setcounter{equation}{0}
\section{Phenomenology for small $\lambda$} \label{sec:pheno}

As we argued in the previous section, if $|\lambda| \ll 1$, the interaction between the two fields will have a small impact on the evolution of fluctuations within the horizon, but it may still have a large impact on super-horizon scales. This fact allows us to deduce the way in which $\lambda$ appears in various observables with the help of simple arguments. In what follows, we discuss a few of them.

\subsection{Computation of the spectra} \label{app:quantization}

In this appendix we analyze the quantization of the coupled system of fields. For convenience, we shall take $\epsilon$, $H$ and $\lambda$ as constants. Let us start by rewriting the quadratic action (\ref{new-action-quadratic}) in terms of canonically normalized variables $u$ and $v$, defined in terms of $\R$ and $\sigma$ as:
\be
u = a \sqrt{2 \epsilon} \R   , \qquad   v = a \sigma .
\ee
By using conformal time $\tau$ to write $a= - 1 / H\tau$ (where $\epsilon \ll 1$ has been used), we find that the action takes the form:
\be
S = \frac{1}{2} \int d^3 x d\tau  \left[ \left( u'  + \frac{\lambda}{\tau}  v \right)^2 +\frac{2}{\tau^2} u^2   + \lambda \frac{2}{\tau^2} u v  -  (\nabla u)^2 +  ( v' )^2 + \frac{2}{\tau^2} v^2     - (\nabla v)^2 \right] ,  \label{action-quadratic-quantum}
\ee
From this result, we infer that the canonical momenta associated to $u$ and $v$ are respectively given by:
\be
\Pi_u = u'  + \frac{\lambda}{\tau}  v  , \qquad  \Pi_v = v' . \label{can-momenta}
\ee
These momenta satisfy the equal time commutation relations, given by
\be
[u(\x, \tau), \Pi_u (\y, \tau)] = i \delta^{(3)} (\x - \y) , \qquad [v(\x, \tau), \Pi_v (\y, \tau)] = i \delta^{(3)} (\x - \y)  ,\label{commut-rel}
\ee
with every other commutation relation vanishing. From (\ref{can-momenta}) we see that the Hamiltonian of the system is given by
\be
H =  \frac{1}{2} \int \! d^3 x \left[ \Pi_u^2 + (\nabla u)^2 - \frac{2}{\tau^2} u^2 + \Pi_v^2 + (\nabla v)^2 - \frac{2}{\tau^2} v^2 - \frac{2 \lambda}{\tau} v   \left(  \Pi_u + \frac{u}{\tau}  \right) \right] .
\ee
We may now split the Hamiltonian into two contributions as $H = H_0 + H_\lambda$, where $H_0$ corresponds to the free Hamiltonian of the system in the case $\lambda = 0$, which corresponds to a a system with two decoupled massless scalar perturbations:
\be
H_0 =  \frac{1}{2} \int \! d^3 x \left[ \Pi_u^2 + (\nabla u)^2 - \frac{2}{\tau^2} u^2 + \Pi_v^2 + (\nabla v)^2 - \frac{2}{\tau^2} v^2 \right] .
\ee
On the other hand, $H_\lambda$ contains the interaction term proportional to $\lambda$:
\be
H_\lambda = - \int \! d^3 x  \frac{\lambda}{\tau} v   \left(  \Pi_u + \frac{u}{\tau} \right)   .
\ee
We may now quantize the system by adopting the interacting picture framework. That is, the quantum fields $u$ and $v$ are expressed as:
\be
u (\x , \tau) = U^\dag (\tau)  u_I (\x , \tau)  U (\tau), \qquad v (\x , \tau) = U^\dag (\tau)  v_I (\x , \tau)  U (\tau), \label{u-v-U}
\ee
where $u_I (\x , \tau)$ and $v_I (\x , \tau)$ are the interaction picture fields, which evolve as quantum fields of the free theory with $\lambda = 0$. Explicitly, they are given by
\be
u_I(\x , \tau) = \frac{1}{(2 \pi)^3} \int d^3 k \, \hat u_I (\k , \tau) \, e^{- i \k \cdot \x }, \qquad 
v_I(\x , \tau) = \frac{1}{(2 \pi)^3} \int d^3 k \, \hat v_I (\k , \tau) \,  e^{- i \k \cdot \x }, 
\ee
with
\be
\hat u_I (\k , \tau) = u_k(\tau) a_{-}(\k) + u_k^*(\tau) a_{-}^\dag(-\k)  , \qquad \hat v_I (\k , \tau) = v_k(\tau) a_{+}(\k) + v_k^*(\tau) a_{+}^\dag(-\k) , \label{Fourier-u-v}
\ee
where the pairs $a_{\pm}(\k)$ and $a_{\pm}^\dag(\k)$ correspond to the creation and annihilation operators satisfying the commutation relations (\ref{comm-a}). The mode functions $u_k(\tau)$ and $v_k(\tau)$ are both given by
\be
u_k(\tau) = v_k(\tau) = \frac{1}{\sqrt{2 k}} \left( 1 - \frac{i}{k \tau} \right) e^{- i k \tau} ,
\ee
which corresponds to the standard expression for a massless mode on a de Sitter space-time with Bunch-Davies initial conditions. On the other hand, $U (\tau)$ is the propagator in the interaction picture, which is given by
\be
U (\tau) = \mathcal T \exp \left\{ - i \int_{-\infty_{+}}^{\tau} \!\!\!\!\!\! d \tau' H_I (\tau')  \right\} ,
\ee
where $\mathcal T$ stands for the time ordering symbol, which, in a given product of operators, it instructs us to put operators evaluated at later times at the left, and operators evaluated at earlier times at the right. In addition, $\infty_{+} = \infty (1 + i \epsilon)$, where $\epsilon$ is a small number introduce to select the correct vacuum in the interacting picture. Given that we are studying a quadratic theory, we may set $\epsilon = 0$ straight away. Notice that the propagator is constructed with $H_I$, which is the interaction picture Hamiltonian given by:
\be
H_I =   - \int \! d^3 x  \frac{\lambda}{\tau} v_I   \left(  \Pi_u^{I} + \frac{u_I}{\tau}  \right) , \qquad \Pi_u^{I} = \frac{d}{d\tau} u_I
\ee
Notice that in the interaction picture, the canonical momenta $\Pi_u^{I}$ coincides with the canonical momenta of the free theory. This ensures that the commutation relations (\ref{commut-rel}) involving the full quantum fields are indeed satisfied. 

Now, we are interested in computing the power spectrum for $\R$ taking into account the leading effects from $\lambda$, assuming that $\lambda \ll 1$. This requires us to expand $u$ in  eq.~(\ref{u-v-U}) up to quadratic order in $\lambda$. In terms of $\hat u (\k , \tau)$ and $\hat v (\k , \tau)$, this expansion has the form:
\bea
\hat u (\k , \tau) &=& \hat u_I (\k , \tau) + i \int_{-\infty }^{\tau} \!\!\!\!\!\!\! d \tau' \left[ H_I (\tau') , \hat u_I (\k , \tau) \right] - \! \int_{-\infty }^{\tau} \!\!\!\!\!\!\! d \tau' \int_{-\infty }^{\tau'} \!\!\!\!\!\!\! d \tau'' \left[ H_I (\tau'') , \left[ H_I (\tau') , \hat u_I (\k , \tau) \right]  \right]   .  \qquad \quad \label{u-u_I} \\
\hat v (\k , \tau) &=& \hat v_I (\k , \tau) + i \int_{-\infty }^{\tau} \!\!\!\!\!\!\! d \tau' \left[ H_I (\tau') , \hat v_I (\k , \tau) \right] - \! \int_{-\infty }^{\tau} \!\!\!\!\!\!\! d \tau' \int_{-\infty }^{\tau'} \!\!\!\!\!\!\! d \tau'' \left[ H_I (\tau'') , \left[ H_I (\tau') , \hat v_I (\k , \tau) \right]  \right]   .  \qquad \quad \label{v-v_I}
\eea
To compute the right hand side of these equations explicitly, it is useful to introduce the following dimensionless functions $A(\tau', \tau)$ and $B(\tau', \tau) $:
\bea
A(\tau', \tau) &\equiv& \frac{i}{k \tau'} \left\{ \cos( k \tau' - k \tau)  + \frac{1}{k \tau} \sin( k \tau' - k \tau)  \right\}, \\
B(\tau', \tau) &\equiv& \frac{i}{k \tau'} \left\{ \left( 1 + \frac{1}{k^2 \tau \tau'} \right) \sin( k \tau' - k \tau)  +    \left( \frac{1}{k \tau'} - \frac{1}{k \tau} \right) \cos( k \tau' - k \tau)  \right\}.
\eea
Then, one finds that the commutators appearing at the right hand side of eqs.~(\ref{u-u_I}) and~(\ref{v-v_I}) are given by:
\bea
\left[ H_I (\tau') , \hat u_I (\k , \tau) \right] &=&  \lambda \, k \, A(\tau', \tau) \,  \hat v_I (\k , \tau') ,\\
\left[ H_I (\tau') , \hat v_I (\k , \tau) \right] &=& \lambda \, B(\tau', \tau) \left(  \frac{d}{d\tau'} \hat u_I (\k , \tau') + \frac{1}{\tau'} \hat u_I (\k , \tau')   \right), \\
\left[ H_I (\tau'') \left[  H_I (\tau') , \hat u_I (\k , \tau) \right] \right] &=&  \lambda^2 \, k \, B(\tau'', \tau') A(\tau', \tau) \left(  \frac{d}{d\tau''} \hat u_I (\k , \tau'') + \frac{1}{\tau''} \hat u_I (\k , \tau'')   \right), \qquad \\
\left[ H_I (\tau'') \left[  H_I (\tau') , \hat v_I (\k , \tau) \right] \right] &=& \lambda^2  \, k^2 \,  B(\tau', \tau)  \frac{\cos(k\tau'' - k \tau') }{k \tau''} \hat v_I (\k , \tau'') .
\eea
Integrating these expressions, and taking the super-horizon limit $|k \tau| \ll 1$, we find that the order-$\lambda$ contributions to eqs.~(\ref{u-u_I}) and~(\ref{v-v_I}) are given by
\bea
&& \int_{-\infty }^{\tau} \!\!\!\!\!\!\! d \tau' \left[ H_I (\tau') , \hat u_I (\k , \tau) \right] = \nn \\ 
&& \qquad \lim_{k \tau' \to - \infty} \frac{ \lambda}{(2 k)^{3/2} \tau} \left( \gamma - 2 + \ln 2 - \frac{i\pi}{2}  + 2 \ln (- k \tau) - \ln ( - k \tau') \right)  a_{+}(\k)  + {\rm h.c.} (-\k), \qquad \\
&& \int_{-\infty }^{\tau} \!\!\!\!\!\!\! d \tau' \left[ H_I (\tau') , \hat v_I (\k , \tau) \right] = \nn \\ 
&& \qquad \lim_{k \tau' \to - \infty} \frac{ \lambda}{(2 k)^{3/2} \tau} \left( \gamma - 2 + \ln 2 - \frac{i\pi}{2}  + \ln ( - k \tau') \right)  a_{-}(\k)  + {\rm h.c.} (-\k) , \qquad 
\eea
where $\gamma$ is the Euler-Mascheroni constant. On the other hand, the order-$\lambda^2$ contributions are:
\bea
&& \! \int_{-\infty }^{\tau} \!\!\!\!\!\!\! d \tau' \int_{-\infty }^{\tau'} \!\!\!\!\!\!\! d \tau'' \left[ H_I (\tau'') , \left[ H_I (\tau') , \hat u_I (\k , \tau) \right]  \right]  = \nn \\
&& \qquad  \lim_{k \tau' \to - \infty} \frac{i \lambda^2}{4 (2 k)^{3/2} \tau} \bigg( - 4 - \frac{\pi^2}{6} + 
\Big[  \gamma - 2 - \frac{i \pi}{2} +  \ln 2 + \ln (- k \tau') \Big]^2  \nn \\ 
&&\qquad + 2 \Big[  \gamma - 2 - \frac{i \pi}{2} +  \ln 2 + \ln (- k \tau') \Big] \Big[  \gamma - 2 - \frac{i \pi}{2} +  \ln 2 - \ln (- k \tau') + 2 \ln (- k \tau) \Big] \bigg) a_{-}(\k) \nn \\ 
&&\qquad 
+  {\rm h.c.} (-\k) ,
\\
&& \! \int_{-\infty }^{\tau} \!\!\!\!\!\!\! d \tau' \int_{-\infty }^{\tau'} \!\!\!\!\!\!\! d \tau'' \left[ H_I (\tau'') , \left[ H_I (\tau') , \hat v_I (\k , \tau) \right]  \right]  = \nn \\
&& \qquad  \lim_{k \tau' \to - \infty} \frac{-i \lambda^2}{4 (2 k)^{3/2} \tau} \bigg( \frac{\pi^2}{2} +  \Big[  \gamma - 2 - \frac{i \pi}{2} +  \ln 2 + \ln (- k \tau') \Big]^2  \bigg) a_{+}(\k) +  {\rm h.c.} (-\k) .
\eea
Notice that there are logarithmic divergences coming from the limit $k \tau' \to - \infty$. These may be traced back to the rotation angle $\theta (\tau) = \frac{\lambda}{2} \ln (- k \tau)$ of the transformation (\ref{Phi}), and therefore, they are expected to cancel out when we compute two point correlation functions, which are independent of the rotation angle. Plugging these results back into eq.~(\ref{u-u_I}), we may compute the two point correlation function $\langle 0 | \hat u (\k , \tau) \hat u^{\dag} (\p , \tau) | 0 \rangle$ up to quadratic order in $\lambda$. The divergent terms cancel out, as expected, and we obtain the finite result
\be
\langle 0 | \hat u (\k , \tau) \hat u^{\dag} (\p , \tau) | 0 \rangle = (2 \pi)^3 \delta(\k - \p) \frac{1}{2 k^3 \tau^2}  \bigg( 1 + \lambda^2 \Big[ A_1 - A_2 \ln (- k \tau) + \ln^2(- k \tau ) \Big]  \bigg) ,
\ee
where:
\bea
A_1 &=& - \frac{\pi^2}{6}  +  (3 - \ln 2) (1 - \ln 2) -  \gamma (4 - \gamma - 2 \ln 2) \simeq - 2.11 , \\
A_2 &=&  4 -2 \gamma - 2 \ln 2 \simeq 1.46. 
\eea
Similarly, we may compute the two point correlation function $\langle 0 | \hat v (\k , \tau) \hat v^{\dag} (\p , \tau) | 0 \rangle$ up to quadratic order in $\lambda$. The result turns out to be independent of $\lambda$:
\be
\langle 0 | \hat v (\k , \tau) \hat v^{\dag} (\p , \tau) | 0 \rangle = (2 \pi)^3 \delta(\k - \p) \frac{1}{2 k^3 \tau^2} .
\ee
Finally, the cross correlation function $\langle 0 | \hat u (\k , \tau) \hat v^{\dag} (\p , \tau) | 0 \rangle$ is found to be given by
\be
\langle 0 | \hat u (\k , \tau) \hat v^{\dag} (\p , \tau) | 0 \rangle = (2 \pi)^3 \delta(\k - \p) \frac{1}{2 k^3 \tau^2}  \lambda \Big[ A_2/2 - \ln (- k \tau)  \Big] .
\ee
To finish, we may use these results to provide expressions for the spectra $\P_\R (k)$. First, notice that $e$-folds are given by $N = \ln(-1/H\tau)$, and that for a given mode $k$ horizon crossing happens at $N_k = \ln(k/H) $. This implies that $\Delta N = N - N_k = - \ln(- k \tau)$. All of this allows us to arrive to the following expression for the power spectrum
\be\label{power-exact}
\P_\R (k) = \frac{H^2}{8 \pi^2 \epsilon}  \bigg( 1 + \lambda^2 \Big[ A_1 + A_2 \Delta N + \Delta N^2 \Big]  \bigg) ,
\ee
which is the desired result.

\setcounter{equation}{0}
\section{Observables} \label{app:observables} 

In this appendix, we deduce the form of some relevant observables, such as the spectral index $n_s$ and the tensor to scalar ratio $r$.

\subsection{Spectral index}

We may now use the results of the previous section to compute the spectral index of the power spectrum for curvature perturbations. We are particularly interested in the spectral index in the limit where $\lambda^2\Delta N^2\gg 1$ remains valid. To do so, it is useful to recall that the combination $\lambda / \sqrt{2 \epsilon}$ appears inside the integral sign, and therefore we can write
\be
\P_\R (t_{\rm end}) =  \frac{H^2}{4 \pi^2 } \bigg( \int_{N_{k}}^{N_{\rm end}} \!\!\!\!\! dN \frac{\lambda}{\sqrt{2 \epsilon }} \bigg)^2 ,
\ee
where $N_k = \ln(k/H) $ is the $e$-fold at which horizon crossing happens (see Appendix~\ref{app:quantization}). Then, the spectral index is given by:
\bea
n_s-1 = \frac{d \ln P_\R}{d \ln k} = 2 \frac{d \ln H}{d \ln k} - 2 \bigg( \int_{N_{k}}^{N_{\rm end}} \!\!\!\!\! dN \frac{\lambda}{\sqrt{2 \epsilon }} \bigg)^{-1}  \frac{\lambda}{\sqrt{2 \epsilon }}  \frac{d N_k}{d \ln k} .
\eea
The important point here is that the running of $\lambda / \sqrt{2\epsilon}$ does not contribute to the spectral index to first order in the slow roll parameters. Next, given that $\lambda / \sqrt{2\epsilon}$ is assumed to evolve slowly, we may write $\int dN \lambda / \sqrt{2\epsilon} \simeq ( \lambda / \sqrt{2\epsilon} ) \Delta N $, and the previous expression is finally well approximated to:
\bea
n_s-1 = - 2 \left( \epsilon  + \frac{1}{\Delta N}  \right) . \label{eq-spectral-index}
\eea
Notice that this expression implies that observations severely restricts the class of models with light fields that can achieve inflation. Both contributions at the right hand side of eq.~(\ref{eq-spectral-index}) have the same sign, and it is not possible to adjust the value of $1 / \Delta N$ so freely. This implies that $\epsilon$ is bounded from above. For instance, if $n_s = 0.96$ and $\Delta N = 60$ then we obtain $\epsilon = 0.0033$.

\subsection{Tensor to scalar ratio}

Finally, let us consider the tensor to scalar ratio $r$. Because tensor modes are not affected by the coupling $\lambda$, it is direct to find that if condition $\lambda^2\Delta N\gg 1$ is satisfied then  $r$ is given by:
\be
r \simeq  \frac{16 \epsilon}{\lambda^2 \Delta N^2}.
\ee
Given that $\epsilon$ must already be very small, as argued in the previous section, we see that $r$ is predicted to be extremely small in this class of models.

\subsection{Relation to single field models with a reduced speed of sound}
Let us note that we might relate the predictions for the two-point function to single-field models of inflation with a reduced speed of sound. In these models, the prediction for the tensor to scalar ratio is given by $r=16\epsilon c_s $, from which we could make the following identification 
\be
c_s=\frac{1}{\lambda^2 \Delta N^2} . \label{eq:c_s}
\ee
In models with a speed of sound different from unity, the spectral index $n_s$ is given by
\be
n_s-1= - (2\epsilon+\eta+s ) \ ,
\ee
where $s=\dot{c_s}/(Hc)$. Using $c_s$ as given in eq.~(\ref{eq:c_s}), we find
\bea
n_s-1 = - 2 \left( \epsilon  + \frac{1}{\Delta N}  \right) \ , \label{eq-spectral-index-cs}
\eea
which is identical to what we found in eq.~(\ref{eq-spectral-index}). At this point, the identification made above should rather be understood as a degeneracy between the two models. A deeper connection could only be endorsed by knowing, for example, whether this identification holds at higher order in perturbation theory.

\end{document}